\documentclass{emulateapj}
\usepackage{graphicx,natbib,color}

\shorttitle{Wave-particle interactions and solar flare electron spectra}
\shortauthors{I. G. Hannah, E. P. Kontar and O. K. Sirenko}

\begin{document}

\title{The effect of wave-particle interactions \\on low energy cutoffs in solar flare electron spectra}

\author{I. G. Hannah\altaffilmark{1}, E. P. Kontar\altaffilmark{1}, and O. K. Sirenko\altaffilmark{2}}
\affil{\altaffilmark{1} Department of Physics and Astronomy, University of Glasgow,
G12 8QQ, UK}

\affil{\altaffilmark{2} Main Astronomical Observatory, Ukrainian Academy of Sciences,
Ukraine}

\begin{abstract}
{Solar flare hard X-ray spectra from RHESSI are normally interpreted in terms of purely
collisional electron beam propagation, ignoring spatial evolution and collective
effects. In this paper we present self-consistent numerical simulations of the spatial
and temporal evolution of an electron beam subject to collisional transport and
beam-driven Langmuir wave turbulence. These wave-particle interactions represent
the background plasma's response to the electron beam propagating from the corona
to chromosphere and occur on a far faster timescale than coulomb collisions. From
these simulations we derive the mean electron flux spectrum, comparable to such
spectra recovered from high resolution hard X-rays observations of solar flares with
RHESSI. We find that a negative spectral index (i.e. a spectrum that increases with
energy), or local minima when including the expected thermal spectral component at
low energies, occurs in the standard thick-target model, when coulomb collisions are
only considered. The inclusion of wave-particle interactions does not produce a local
minimum, maintaining a positive spectral index. These simulations are a step towards
a more complete treatment of electron transport in solar flares and suggest that a flat
spectrum (spectral index of 0 to 1) down to thermal energies maybe a better
approximation instead of a sharp cut-off in the injected electron spectrum.}
\end{abstract}

\keywords{Sun: flares - Sun: X-rays, gamma rays - Sun: activity -Sun: particle emission}

\section{Introduction}

Hard X-ray emission has long been used as the prime diagnostic tool to study particle
acceleration and energy release in solar flares. From these X-ray observations the
mean electron flux spectrum (e.g. averaged over the X-ray emitting volume, see
\citet{Brown_etal2003} for details) can be determined either through forward fitting
\citep{Holman_etal2003} or more advanced inversion techniques
\citep{Piana_etal2003,Kontar_etal2004,Brown_etal2006}. At higher energies, typically
above 10-20~keV, the observed hard X-ray spectrum is considered to be due to an
accelerated population of electrons being stopped by the dense chromosphere via
Coulomb collisions \citep{1971SoPh...18..489B}. The spectrum below 10-20~keV
normally originates from thermal coronal sources with temperatures of 10s MK
\cite[e.g.][]{KruckerLin2008}.

The Reuven Ramaty High Energy Solar Spectrometer (RHESSI) provides high resolution
HXR spectra of solar flares \citep{2002SoPh..210....3L}, greatly improving on previous
measurements \citep{JohnsLin1992}. This high energy resolution spectra has allowed,
for the first time, to scrutinize the X-ray and electron spectra in search for a
non-powerlaw features, revealing vital clues about electron acceleration and
transport. In some RHESSI flares, the recovered mean electron flux spectrum
demonstrates a local minima or \emph{dip} between the non-thermal and the thermal
components instead of a smooth transition
\citep{Piana_etal2003,Holman_etal2003,Kasparova_etal2005,Sui_etal2007,Kontar_etal2008}.
The presence of the dip, as a real physical feature of the electron spectra, has been
questioned as in many cases it can be attributed to photospheric albedo, e.g.
Compton back-scattered X-rays \citep{Kasparova_etal2005,Kontar_etal2008}.
However a few events have been found in which after isotropic albedo correction
\citep{Kontar_etal2008}, the X-ray spectrum is still relatively flat, so they could be
fitted with a thick-target model single power-law spectrum with a low energy cutoff
\citep{Sui_etal2007}. In these flares a dip was not directly observed in the mean
electron spectrum, but instead inferred from forward fitting a model with low energy
cutoff to the X-ray spectrum. This model has a thermal component at low energies
and at higher energies purely collisional thick-target model of a single power-law of
accelerated electrons above a cutoff. In this thick-target scenario
\citep{1971SoPh...18..489B,Holman2003}, the dip in the mean electron spectrum
originates from a positive slope at low energies developing below the cutoff as the
accelerated electrons propagate from a coronal acceleration site downwards to the
chromosphere, having Coulomb collisions with the background plasma. If the dip is
real it provides important insights into flare energetics since the energy in
accelerated electrons is strongly dependent on the low energy cutoff.

For any reasonable X-ray producing flare, non-collisional beam-plasma interaction is
much faster than that via Coulomb collisions
\citep{ZheleznyakovZaitsev1970,Karlick2009}. Such processes are inferred to occur in
downward propagating electron beams from radio observations of reverse slope drift
burst in flares \citep[e.g.][]{1997A&A...320..612K,1997ApJ...480..825A}. Although
generation and escape of electromagnetic radiation from Langmuir waves in a flaring
plasma is not well understood. The role of wave-particle interactions in solar flares
assuming stationary, time-independent injection of electrons have been considered
analytically and numerically
\citep{1984ApJ...279..882E,hamilton1987,McClements1987}.
\citet{1984ApJ...279..882E} have argued that the conditionally created distribution
should be constantly being flattened by quasi-linear relaxation, while
\citet{hamilton1987} and \citet{McClements1987} suggest that although the
wave-particle interactions have an important effect, the change of the electron
spectra under stationary conditions should be minor. However, in a more realistic
flare conditions the injection (acceleration) of electrons is likely to be highly
intermittent \citep{TsiklauriHaruki2008} with a number of short duration pulses is
often observed
\citep{Kiplinger_etal1984,1994SoPh..153..403F,Aschwanden_etal1998}, so the
time-dependent solution of particle transport equations accounting for wave-particle
interactions should be considered. Additionally, the previous studies did not consider
the spatial and temporal evolution of the beam from the coronal source down into
chromosphere - a crucial aspect when considering the propagation of an electron
beam in comparison to X-ray observations.

\begin{figure}\centering
\plotone{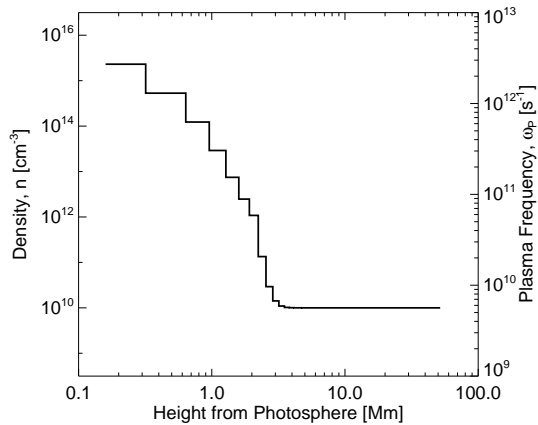}\\
\caption{\label{fig:density} The background plasma density $n(x)$
as a function of the height above the photosphere. Also shown is the
corresponding plasma frequency for each of these densities.}
\end{figure}

The quasi-linear relaxation process of Langmuir waves has been considered in higher
velocity dimensions (other than the parallel component considered here)
\citep[e.g.][]{1980SvJPP...6Q.422C} but it has only been recently that the 2D system
has been fully numerically solved \citep{2008PPCF...50h5011Z}. Even then the
evolution was considered in a spatially independent manner. In these studies it was
found that the parallel component (1D) is the fastest processes and likely to dominate
the electron transport.

In this letter we take a step towards a more complete treatment of electron transport
in solar flares by including the spatial evolution of beam-driven Langmuir wave
turbulence. We numerically study the system self-consistently, simulating the
propagation of an electron beam from the coronal acceleration site down to the
chromosphere, considering the truncated power-law spectrum frequently used for
data interpretation \citep[e.g.][]{Holman2003,Sui_etal2007}, to investigate the
evolution of the mean electron flux spectrum below this cutoff. We demonstrate that
the positive slope of the mean electron flux is not present, when the response of the
background plasma via Langmuir waves to the propagating electron beam is taken
into account. We also show that the injected electron spectrum flattens to a
decreasing distribution due to collective interaction with plasma even for weak flares
\citep[e.g.][]{hannah2008}. Furthermore, we suggest that a flat spectrum (with
spectral index 0 to 1) down to thermal energies maybe a better approximation as
opposed to the sharp cut-off in the injected electron spectrum.

\section{Particle transport and wave-particle interaction}\label{sec:model}

Following the standard model approach for interpreting solar flare hard X-ray spectra,
we assume the electron flux spectrum of injected (flare accelerated) electrons is a
power law, $F(E)\sim E^{-\delta}$, [electrons cm$^{-2}$ s$^{-1}$ keV$^{-1}$] down to
some energy $E_C$, typically 10-20 keV. The initial 1D electron distribution function
(accelerated electron population) subsequently is also a power-law in velocity
$f(v)=F(E)/m$, above a low energy cutoff $v_\mathrm{C}$ with spectral index
$\alpha=2\delta$. For our simulations we consider such an initial electron distribution
which is also spatially a gaussian of characteristic size $d$

\begin{equation}\label{eq:init}
 f(v,x,t=0)=n_\mathrm{0}\frac{(1-\alpha)}{v_\mathrm{C}
}\left(\frac{v}{v_\mathrm{C}}\right)^{-\alpha} \exp{\left(-\frac{x^2}{d^2}
\right)}
 \end{equation}

\noindent normalised by the beam density $n_\mathrm{0}$.

\begin{figure*}\centering
\includegraphics[width=1.5\columnwidth]{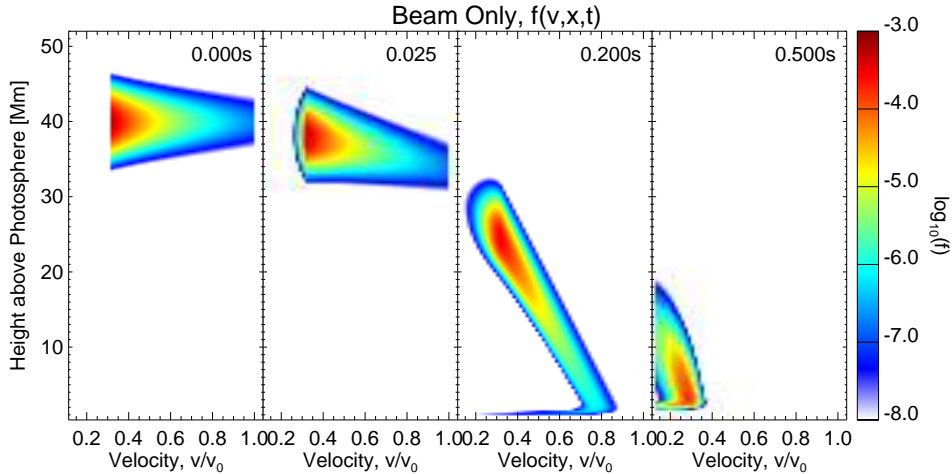}\\
\caption{\label{fig:fb}
The evolution of the electron distribution $f(v,x,t)$ (with time increasing from
left to right) for the simulation only following the progression of an electron
beam subject to Coulomb collisions, i.e. the standard thick-target model
\citep{1971SoPh...18..489B}. (An animation of Figure \ref{fig:fb} and
\ref{fig:fbw} is available in the online journal.)}
\end{figure*}

To self-consistently follow the temporal and spatial evolution of an electron beam
from a coronal acceleration site, including the response of the thermal background
plasma in the form of Langmuir waves, we use the 1D equations of quasi-linear
relaxation \citep{VedenovVelikhov1963,DrummondPines1964,
1969JETP...30..131R,hamilton1987,2001SoPh..202..131K}

\begin{equation}\label{eq:f}
  \frac{\partial f}{\partial t} +v\frac{\partial f}{\partial x}=
\frac{4\pi^2
e^2}{m^2}
  \frac{\partial}{\partial v}\left( \frac{W}{v}\frac{\partial f}{\partial
v}\right)+\gamma_\mathrm{C_F}\frac{\partial}{\partial
v}\left(\frac{f}{v^2}\right)
\end{equation}
\begin{equation}\label{eq:w}
  \frac{\partial W}{\partial t}+ \frac{3v_\mathrm{T}^2}{v}\frac{\partial
W}{\partial
x} =   \left( \frac{\pi \omega_\mathrm{p}}{n}v^2  \frac{\partial f}{\partial
v}-\gamma_\mathrm{C_W}-2\gamma_\mathrm{L} \right)W+Sf
\end{equation}

\noindent where $f(v,x,t)$ is the electron distribution function [electrons cm$^{-4}$
s], $W(v,x,t)$ is the spectral energy density [erg cm$^{-2}$], $k$ is the wave number
of a Langmuir wave, $n$ is the background plasma density and
$\omega_\mathrm{p}^2=4\pi n e^2/m$ is the local plasma frequency. The first
component on the righthand side of equations (\ref{eq:f}) and (\ref{eq:w}) are the
quasilinear terms that describes the resonant interaction between the electrons and
Langmuir waves, $\omega_\mathrm{p}=kv$. Also included are the Coulomb collision
dampening rate for both the electrons $\gamma_\mathrm{C_F}=4\pi e^4
n\ln{\Lambda}/m^2$ \citep{1978ApJ...224..241E} and waves
$\gamma_\mathrm{C_W}=\pi e^4n \ln{\Lambda}/(m^2 v_\mathrm{T}^3)$
\citep{1980panp.book.....M}. Where $\ln{\Lambda}=\ln{(8\times 10^6n^{-1/2}T)}$ is
the Coulomb logarithm, $T$ is the temperature of the background plasma and
$v_\mathrm{T}=\sqrt{k_\mathrm{B}T/m}$ is the velocity of a thermal electron,
$k_\mathrm{B}$ is the Boltzmann constant. Also included in equation (\ref{eq:w}) is
the Landau dampening rate
$\gamma_\mathrm{L}=\sqrt{\pi/8}\omega_\mathrm{p}\left(v/v_\mathrm{T}
\right)^3\exp{ \left(-v^2/2v_\mathrm{T}^2\right ) }$ \citep{1981phki.book.....L} and
the spontaneous emission $S=\omega_\mathrm{p}^3m
v\ln{\left(v/v_\mathrm{T}\right)}/(4\pi n)$
\citep{1980panp.book.....M,Tsytovich1995,hamilton1987}.

In the simulations, we take $\alpha=7$, or $\delta=3.5$ from a cutoff of
$v_\mathrm{C}=7.26\times10^9$ cm s$^{-1}$, or $E_\mathrm{C}=15$ keV, up to
maximum of $v_\mathrm{0}=2.4\times10^{10}$ cm s$^{-1}$. Our simulation extends
in velocity space from $v=7v_\mathrm{T}=2.73\times10^{9}$  cm s$^{-1}$ (taking
$T=1$MK) to $v=2.5\times10^{10}$ cm s$^{-1}$. The initial spatial scale of the beam
is $d=2\times10^8$~cm with density of $n_\mathrm{0}=10^6$ cm $^{-3}$. The initial
number of electrons in this simulation is $N\approx 23 n_\mathrm{0}d^3=10^{32}$
electrons, which is an approximation as we only have one spatial dimension to
estimate the volume from, taking this as the $FWHM=2\sqrt{2\ln{2}}d$, which is
typically measured is X-ray imaging observations
\citep[e.g.][]{Hurford_etal2002,Kontar_etal2008footp}. Here we have used a modest
number of electrons, similar to that found in a small flare, or microflare
\citep{hannah2008,2008A&A...481L..45H}, a typical A or B-class GOES flare. Since the
rate of wave-particle interactions is proportional to electron beam density, the effects
of wave-particle interaction will be present to a far greater extent in larger flares.

\begin{figure*}\centering
\includegraphics[width=1.5\columnwidth]{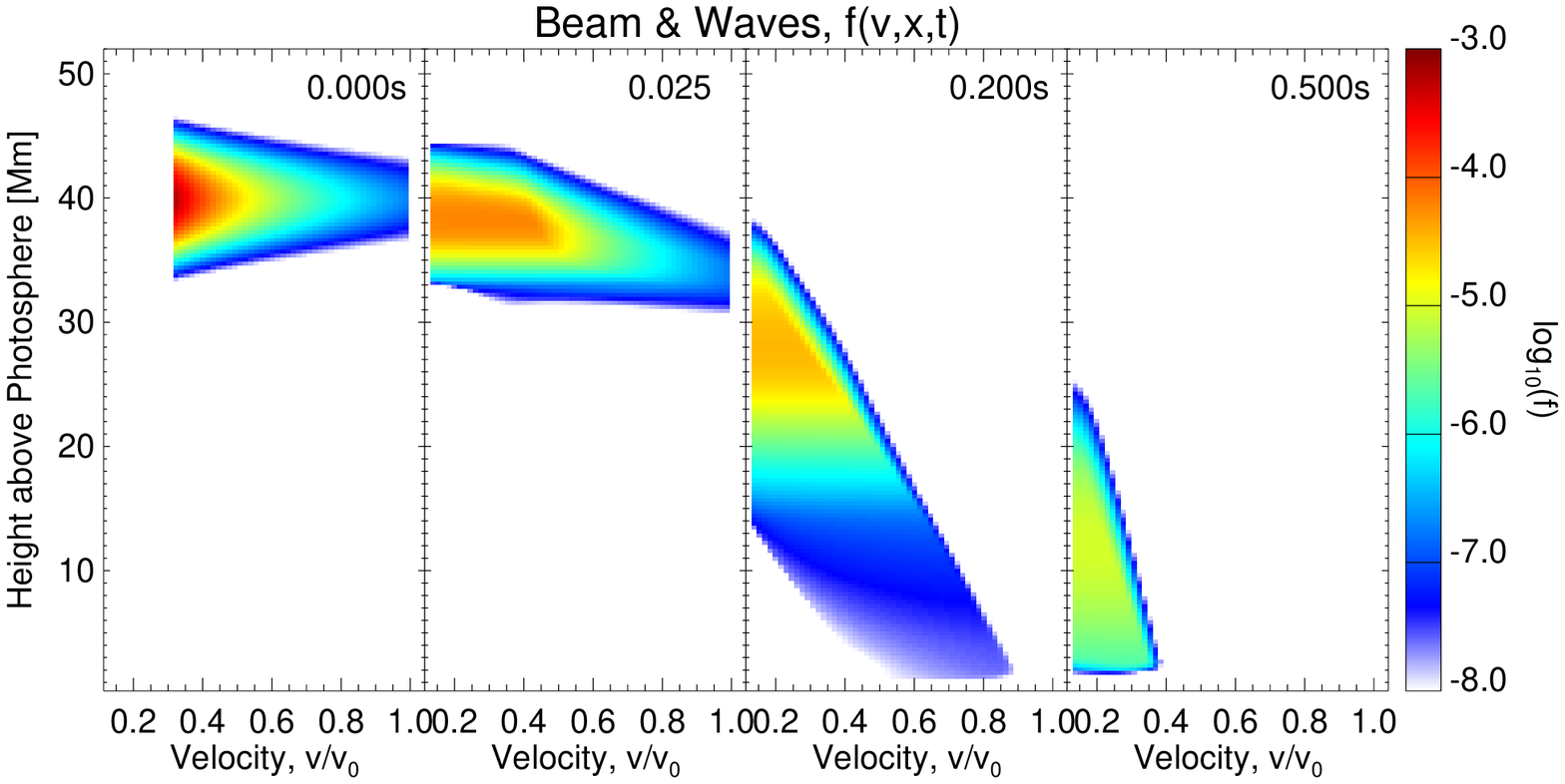}\\
	\includegraphics[width=1.5\columnwidth]{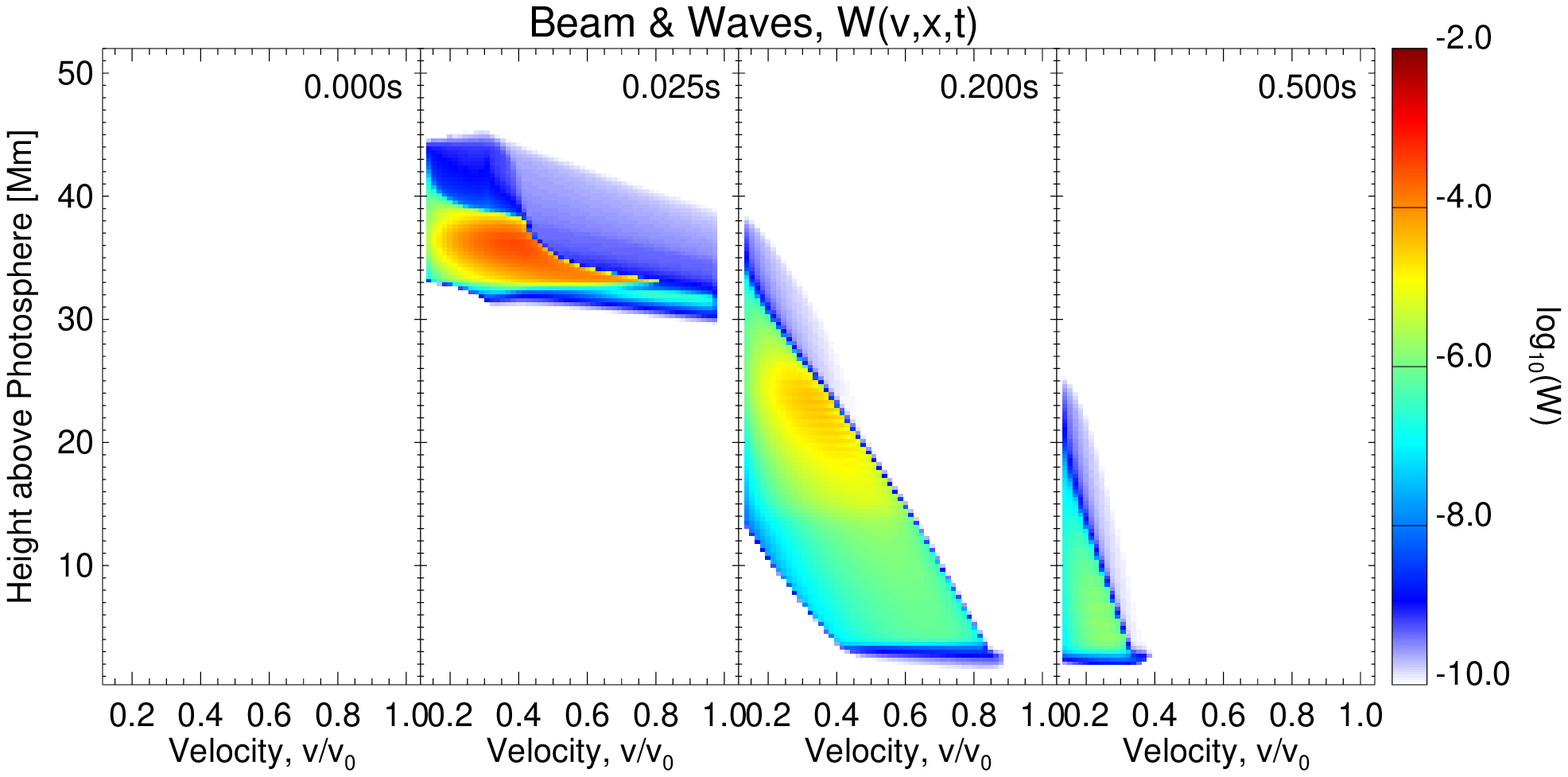}\\
\caption{\label{fig:fbw}The electron distribution $f(v,x,t)$, top row, and wave
spectral
energy density $W(v,x,t)$, bottom row, with time increasing from left panel to
right
for the simulation following the progression of an electron beam including
Langmuir
wave response from the background plasma. $W$ is shown against electron velocity
$v$ instead of wave number $k$ since $k=\omega_p/v$ and to allow easy comparison
between the two rows. (An animation of Figure \ref{fig:fb} and \ref{fig:fbw} is
available in the online journal.)}
\end{figure*}

We approximate the background plasma density $n$ assuming a constant of
$10^{10}$cm$^{-3}$ at coronal heights, with a sharp density increase at the
chromosphere level, with further steady hydrostatic increase towards the photosphere
\citep{Aschwanden_etal2002}, see Figure \ref{fig:density} The initial electron beam is
spatially centred at $h_\mathrm{0}=4\times 10^{9}$~cm, see Figures \ref{fig:fb} and
\ref{fig:fbw} for details.

\begin{figure}\centering
\plotone{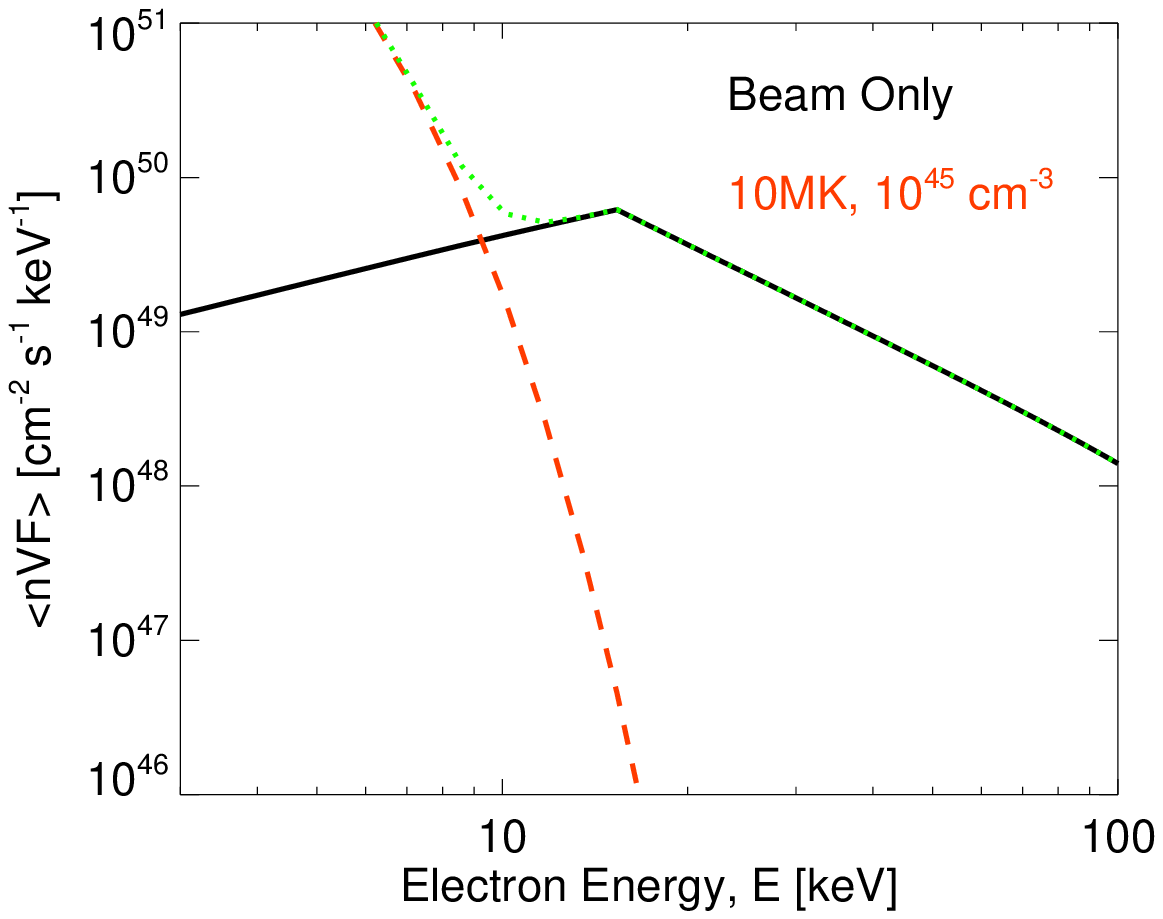}\\
\plotone{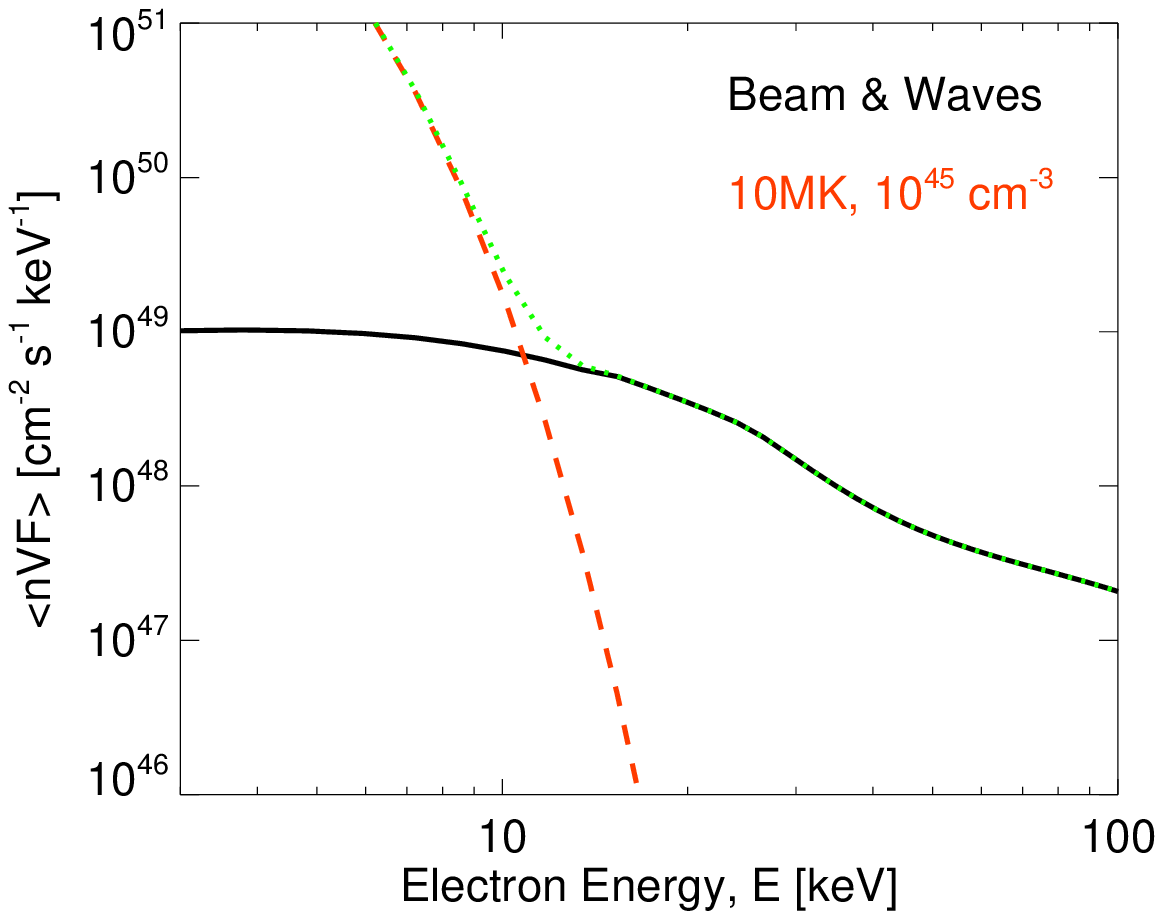}
\caption{\label{fig:nfv} The mean electron flux spectrum
$<nVF>$ for the simulation of an electron beam with only
collisional dampening (top) and for the electron beam with the
generation of Langmuir waves (bottom). These
spatially integrated spectra are averaged over the
1sec duration of the simulations. The dashed lines are a thermal model with
$T=10$ MK and $EM=10^{45}$ cm$^{-3}$. The dotted lines are the total spectrum.}
\end{figure}

The equations (\ref{eq:f}) and (\ref{eq:w}) are solved numerically using a finite
difference method as described by \citet{2001CoPhC.138..222K}. This is over a grid of
60 points in velocity space and 160 in position space. The fastest process here is the
quasilinear relaxation, occurring on a timescale of $\tau_\mathrm{Q}\approx
n_\mathrm{0}n/ \omega_\mathrm{p}\sim 2\times10^{-5} \sqrt{n}/n_\mathrm{0}$
seconds. Therefore we numerically solve equations (\ref{eq:f}) and (\ref{eq:w}) using
a time-step at least an order of magnitude smaller. The initial spectral energy density
is taken to be the thermal background which has reached a steady-state through
Coulomb collisions and wave-particle emission/absorption. These simulations are ran
for 1 second in simulation time, enough time for all of the electrons to reach the
highest density region, lose energy and then leave the simulation grid, joining the
thermal electrons.

\subsection{Beam Coulomb Collisions}\label{sec:bonly}

We start by simulating the propagation of the electron beam in the absence of waves,
with only Coulomb collisions acting on the electrons, following the standard
thick-target model \citep{1971SoPh...18..489B}. Namely, only solving equation
(\ref{eq:f}) and ignoring the first term on the righthand side, the wave-particle
interactions. The resulting electron distribution $f(v,x)$ for various times during the
simulation is shown in Figure \ref{fig:fb}. The electrons with the highest velocities
move quickly to lower heights where they encounter the sharply increasing
background plasma density (chromosphere) below about 3 Mm. Here the Coulomb
collisions quickly cause the electrons to lose energy and eventually have velocities
outside of the simulation grid. At lower energies the sharp initial low energy cutoff is
smoothed out through Coulomb collisions reducing the electrons velocity. The time
averaged mean electron flux spectrum of purely collisional transport is shown in
Figure \ref{fig:nfv}. This spatially integrated mean electron flux spectrum $<nVF>$ or
$\overline{n}V\overline{F}$ is related to the simulated electron distribution $f(v,x,t)$
as $\overline{n}V\overline{F}(E,t)=A \sum \left[n(x)f(v,x,t)/m_\mathrm{e}\right] dx $
\noindent where $A$ is the cross-sectional area of the beam. We take this to be
$FWHM^2 \approx5.5d^2$ given our 1D simulation. The positive slope at low
energies is clearly visible with the expected decreasing power-law above roughly the
original low energy cutoff (Figure \ref{fig:nfv} left). Overplotted is an example model
thermal spectrum to demonstrate how the local minima (dip) appears in the spectrum.
A typical flare temperature of 10 MK is used, with a modest emission measure of
$10^{45}$ cm$^{-3}$ to match the observations of the small flare we have simulated.
This is a standard thick-target spectrum, which is used to fit and interpret X-ray
spectrum.

\subsection{Beam-Driven Langmuir Waves}\label{sec:bw}

\begin{figure}\centering
\includegraphics[width=0.9\columnwidth]{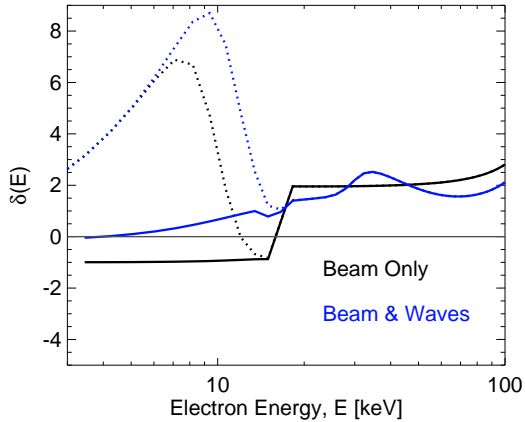}\\
\caption{
\label{fig:deriv}The spectral index $\delta (E)=-d\log(<nVF>)/d\log(E)$ of the
mean electron spectrum for the simulation with the electron beam only subject to
Coulomb collisions (black line) and with the beam-driven Langmuir waves (blue
line). The dashed line indicates the total model spectra (including thermal
component) demonstrating a local minima in the beam only case.} \end{figure}

We now follow the beam propagating with Langmuir waves generated by the
background plasma in response to the beam, numerically solving both equations
(\ref{eq:f}) and (\ref{eq:w}). The resulting electron and spectral wave density
distributions are shown in Figure \ref{fig:fbw} as a function of $v$ and $x$ for various
times during the simulation. The electron distribution function $f(v,x,t)$ quickly
flattens to form a plateau-like distribution expected for beam-plasma interaction via
plasma waves \citep[e.g.][]{ZheleznyakovZaitsev1970}. The electrons together with
the waves move down and eventually end up in the dense regions of the atmosphere,
where the transport becomes dominated by collisions. The motion of the plasma
turbulence (Figure \ref{fig:fbw}) is not due to the group motion of waves, which is
negligibly small $\partial\omega/\partial k =3v_{Te}^2/v \ll v $, but appears because
the Langmuir waves are locally generated and efficiently reabsorbed by the beam
itself or collisionally by the surrounding plasma. The wave-particle interactions clearly
changes the overall shape of spatially integrated electron flux spectrum, as shown in
the right panel of  Figure \ref{fig:nfv}. Crucially, no positive slope is created in the
non-thermal spectrum below the initial low energy cutoff, resulting in no dip in the
overall model spectrum. This can be further seen when we consider the spectral index
$\delta$ of the mean electron flux spectrum as a function of energy in Figure
\ref{fig:deriv}. Where as the beam only simulation produces a brief energy range
where $\delta <0$, it is always positive in the beam and wave case. The slightly
reduced level of the mean electron flux spectrum also suggests that the generation of
Langmuir waves leads to additional energy losses by the beam (into heating the
background plasma) and higher number of energetic electrons will be required to
explain the same X-ray spectra.

\section{Discussion \& Conclusions}

Considering both the temporal and spatial evolution, the mean electron flux spectrum
is very sensitive to the generation of waves. The influence of wave-particle
interactions is seen to flatten the spectral index of the electron spectrum $\delta
(E)=d\log(\overline{n}V\overline{F})/d\log(E)$ below the break but also up to where
the beam-plasma interaction time is faster than the electron cloud time size. The
artificially introduced low-energy cutoff in the injected electron spectrum disappears
not only in the local electrons distribution function  but also in the spatially integrated
electron spectrum once wave-particle interactions is taken into account. The
character of beam propagation is close to the simulation results of beam transport
along open field lines, where collisions are normally ignored
\citep{TakakuraShibahashi1976,MagelssenSmith1977,Melnik_etal1999,Li_etal2008}.
However, there are noticeable differences. In the simplistic treatment of spatially
uniform beam, when all but quasilinear terms are ignored
\citep[e.g.][]{VedenovVelikhov1963}, the generation of waves leads to an exact
plateau distribution with $\delta (E)\approx0$. Our simulations show that the spectral
index, $\delta (E)$ is more than zero, which is the result of collisions. Similarly,
\citet{KontarReid2009} show that the spatially integrated spectrum of particles will
not deviate from a initial power-law, but only when processes leading to absorption of
waves or removal of waves out of resonance are included.

The simulations in this letter are for typical microflare parameters
\citep{hannah2008} and given that the wave emission scales with the electron
number density we would expect wave-particle interactions to have more significant,
effect in large flares. Large flares could constitute multiple intermittent bursts of
accelerated electrons directed along possibly different magnetic field lines. The fast
time variations in hard X-ray lightcurves \citep{Kiplinger_etal1984} indirectly support
this idea. Numerical simulations of reconnection suggest ``bursty'' electron
acceleration \citep{TsiklauriHaruki2008} and spatially fragmented electron
acceleration \citep{BianBrowning2008} which could result in electron propagation
along the different lines. The footpoint motion often seen in solar flares
\citep{Krucker_etal2003} also suggest that the electrons are consecutively injected
onto field lines. In this scenario an ensemble of our simulations, multiple micro-beam
injections, would lead to beam densities comparable to a large flare.

The convergence of the magnetic field at chromospheric heights
\citep[e.g.][]{2008A&A...489L..57K} has been ignored in this work, however in
our
simulations the overall evolution of the energetic particles in the top part of a
relatively dense loop, $10^{10}$~cm$^{-3}$, is dominated by wave-particle
interactions where the field is not converging. It is only in the denser chromosphere,
where the field is likely to converge, that the collisions become dominant. The very
fast flattening of the powerlaw distribution's low energy cutoff by the wave-particle
interactions suggests that it is unlikely that such a cutoff could develop and is
therefore an unwise initial distribution for any model of coronally accelerated
electrons. The non-thermal distribution flattening at low energies as it transitions into
the thermal distribution seems to be a realistic model.  Given how strongly the total
energy in the accelerated elections depends on a cutoff, or its behaviour at low
energies approaching the thermal distribution
\citep{2003ApJ...595L.119E,2005A&A...438.1107G}, this transition needs further
study.

The work presented here is a step towards a more complete treatment of electron
transport in solar flares, moving from the standard thick target model (1D velocity
with collisions only) to beam-driven Langmuir turbulence (1D velocity, 1D space,
collisions and wave-particle interactions). In future work we need to extend this to
consider 2 or higher dimensions of velocity space, the changing magnetic field,
electron scattering as well as the transition between the accelerated beam and
thermal distribution.

\acknowledgments

We would like to thank the referee for their constructive comments. This work is
supported by a STFC rolling grant (IGH, EPK) and STFC Advanced Fellowship (EPK).
Financial support by the Royal Society grant (RG090411), The Royal Society Short
Incoming Visit Grant (IV0871184), and by the European Commission through the
SOLAIRE Network (MTRN-CT-2006-035484) is gratefully acknowledged.


\clearpage

\end{document}